\documentclass[prl,amsfonts,english,showpacs,twocolumn,floatfix]{revtex4}
\usepackage[T1]{fontenc}
\usepackage[latin1]{inputenc}
\usepackage{amsmath}
\usepackage{graphicx}
\usepackage{amssymb}

\makeatletter
\usepackage{babel}
\makeatother
\begin{document}

\title{Long-range correlations and ensemble inequivalence in a generalized ABC model}

\author{A. Lederhendler and D. Mukamel}

\affiliation{Department of Physics of Complex Systems, Weizmann Institute of Science,
76100 Rehovot, Israel}

\date{\today}

\begin{abstract}
A generalization of the ABC model, a one-dimensional model of a
driven system of three particle species with local dynamics, is
introduced, in which the model evolves under either (i) density-conserving 
or (ii) nonconserving dynamics. For equal average
densities of the three species, both dynamical models are
demonstrated to exhibit detailed balance with respect to a
Hamiltonian with long-range interactions. The model is found to
exhibit two distinct phase diagrams, corresponding to the canonical
(density-conserving) and grand canonical (density nonconserving) ensembles,
as expected in long-range interacting systems. The implication of
this result to nonequilibrium steady states, such as those of the ABC
model with unequal average densities, are briefly discussed.

\end{abstract}

\maketitle

In many cases nonequilibrium systems driven by an external field,
such as the temperature gradient or electric field, reach a
nonequilibrium steady state which is characterized by long-range
correlations \cite{Spohn1983}. In some cases such long-range
correlations lead to phenomena like the emergence of long-range
order and spontaneous symmetry breaking, even when the dynamics is
local and stochastic \cite{Mukamel2000}. This is particularly intriguing in one-dimensional 
models, where such phenomena do not exist under
equilibrium conditions for systems with short-range interactions at
finite temperatures.

A model which is often discussed in this context is the ABC model
\cite{Evans1998, Clincy2003, Ayyer2009}. This is a one-dimensional
three species model of a driven system with {\it local dynamics}
which exhibits phase separation at nonvanishing densities of the
three species. It has been shown that at a particular point of its
phase space, namely, for equal densities of the three species, the
model obeys detailed balance, and the steady-state distribution can
be expressed in terms of an effective Hamiltonian with {\it
long-range} interactions. Thus, although the dynamics is local and
stochastic, the long-range nature of the resulting effective
interactions leads to phase separation and long-range order.

Equilibrium systems with short-range interactions typically evolve
into a unique state which is independent of their dynamics. Thus,
for instance, the Ising models with magnetization-conserving
(Kawasaki) dynamics and nonconserving (Glauber) dynamics result in
the same equilibrium state. This equivalence of ensembles is broken
in systems with long-range interactions, where the two-body
potential decays at large distance with a power smaller than the
spatial dimension. Such systems are nonadditive and their
thermodynamic functions are superextensive with respect to their
size. As a result, they exhibit unusual phenomena such as ensemble
inequivalence, negative specific heat, and slow relaxation dynamics
(see, for example, \cite{Thirring1970,Mukamel2010,Campa2009}). The ABC
model is expected to share some of these properties at the equal
densities point, where it behaves as an effective long-range
interacting system. A different mechanism for ensemble inequivalence
has been discussed within the context of the zero-range process,
whereby long-range order emerges in one dimension due to the noncompactness 
of the local order parameter \cite{Grosskinsky2008}.

In this Letter, we generalize the ABC model to include vacancies and
nonconserving processes. This allows us to compare the phase
diagram of the conserving model with that of the
nonconserving model. We show that for equal densities both dynamics
obey detailed balance with respect to a Hamiltonian with long-range
interactions. Under conserving dynamics, where the
nonconserving processes are excluded, the resulting steady state is
that of the canonical ensemble. The nonconserving dynamics
correspond to the grand canonical ensemble. Evaluating the phase
diagram of the generalized model, we find that, as is common in
systems with long-range interactions, the canonical and the grand
canonical ensembles are inequivalent, yielding different phase
diagrams.

Small deviations from the equal densities condition are not expected to
significantly alter the steady state. Thus, a detailed study of the model
with equal densities may serve as a guideline for investigation of the
general case with nonequal densities, where detailed balance is not
satisfied and an effective energy cannot be defined. This may shed
some light on the mechanisms leading to long-range phenomena in
nonequilibrium steady states.

We begin by outlining the main features of the ABC model on a ring
\cite{Evans1998}. The model consists of three species of particles,
labeled $A,\, B$, and $C$, which occupy the sites of a periodic
lattice of length $L$. The number of particles of each type is given
by $N_{A},\, N_{B}$, and $N_{C}$ with $N_{A}+N_{B}+N_{C}=L$. The
model evolves by local, random sequential dynamics whereby two
neighboring particles are exchanged clockwise with the following
rates:
\begin{equation}
AB\overset{q}{\underset{1}{\rightleftarrows}}BA\,;\,
BC\overset{q}{\underset{1}{\rightleftarrows}}CB\,;\,
CA\overset{q}{\underset{1}{\rightleftarrows}}AC.\label{eq:ABCdynamics}
\end{equation}
For $q=1$ the dynamics is symmetric and the system relaxes to a
homogeneous state, in which all particles are uniformly distributed,
regardless of type. For $q \ne 1$ the system reaches a
nonequilibrium steady state in which the particles phase separate
into three domains. The domains are arranged, clockwise, in the
order $AA\ldots ABB\ldots BCC\ldots C$ for $q<1$ and
counterclockwise for $q>1$. We will assume $q<1$ for the rest of the
Letter. It has been shown that at equal densities, $N_{A}=N_{B}=N_{C}=L/3$,
the dynamics obeys detailed balance with respect to the Hamiltonian
\begin{equation}
\mathcal{H}\left(\left\{ X_{i}\right\} \right)=\sum_{i=1}^{L-1}
\sum_{j=1}^{L-i}\left(A_{i}C_{i+j}+B_{i}A_{i+j}+C_{i}B_{i+j}\right).\label{eq:H_ABC}
\end{equation}
Here $\left\{ X_{i}\right\} =\left\{ A_{i},B_{i},C_{i}\right\} $,
\begin{equation}
A_{i}=\left\{ \begin{array}{cc}
1 & {\rm if\, site\,}i{\rm \, is\, occupied\, by\, an\,}A{\rm \, particle}\\
0 & {\rm otherwise,}\end{array}\right.
\end{equation}
and similarly for $B_i$ and $C_i$. The change in $\mathcal{H}$ due
to the exchange of any pair of neighboring particles as described in
\eqref{eq:ABCdynamics} is $\Delta\mathcal{H}=\pm1$, where, for example,
the exchange of $AB$ to $BA$ raises the energy by $1$. Detailed balance
is thus maintained with respect to the steady-state distribution
$P\left(\left\{ X_{i}\right\}\right)=q^{\mathcal{H}\left(\left\{
X_{i}\right\} \right)}/Z_{L}$, with the partition sum
$Z_{L}=\sum_{\left\{ X_{i}\right\} }q^{\mathcal{H}\left(\left\{
X_{i}\right\} \right)}$. While the microscopic dynamics
\eqref{eq:ABCdynamics} is strictly local, the interactions in
$\mathcal{H}$ are long-ranged (mean-field like).

It has been demonstrated that, by considering an L-dependent $q$
which approaches $1$ sufficiently fast at large L, the $q=1$
homogeneous state may be realized \cite{Clincy2003}. For
$q=\exp\left(-\beta/L\right)$, the model exhibits a phase transition
from a homogeneous state for $\beta<\beta_{c}$ to a phase-separated
state for $\beta>\beta_{c}$, with $\beta_{c}=2\pi\sqrt{3}$ . The
parameter $\beta$ serves as an inverse temperature.

We now generalize the model, allowing us to compare conserving and
nonconserving dynamics. As a first step we introduce vacancies into
the model. Thus, each site may be occupied either by one of the
three species or by a vacancy $0$. Hence, $N\equiv
N_{A}+N_{B}+N_{C}\leq L$. The dynamics is such that vacancies are
neutral, so that a particle of any species may hop to the left or
right into an empty site with equal probability. The following rule
is added to the exchange rules Eq. \eqref{eq:ABCdynamics}:
\begin{equation}
X0\overset{1}{\underset{1}{\rightleftarrows}}0X\, , \;
X=A,B,C.\label{eq:vacancyexchange}
\end{equation}

Next, we add nonconserving processes, whereby particles are allowed
to leave and enter the system in ordered groups of three neighboring
particles:
\begin{equation} ABC\overset{pq^{3\mu
L}}{\underset{p}{\rightleftarrows}}000,\label{eq:addremoverates}
\end{equation}
where $\mu$ is a chemical potential, taken to be equal for all three
species, and $p$ is a parameter whose value does not affect the
steady state in the case where detailed balance is satisfied. This
particular choice of the nonconserving process maintains equal
densities
whenever the initial configuration satisfies this condition.

Focusing on the case of equal densities, we consider two alternative dynamics:
(i) density-conserving dynamics where the evolution takes place by
the processes \eqref{eq:ABCdynamics} and \eqref{eq:vacancyexchange}
and (ii) nonconserving dynamics where all processes
\eqref{eq:ABCdynamics}, \eqref{eq:vacancyexchange} and
\eqref{eq:addremoverates} are allowed. Under both dynamics, detailed
balance is satisfied with respect to the following Hamiltonian:
\begin{equation}
\tilde{\mathcal{H}}\left(\left\{ X_{i}\right\} \right)= \mathcal{H}\left(\left\{ X_{i}\right\} \right) -\frac{N\left(N-1\right)}{6}-\mu NL.\label{eq:H_ABCgen}
\end{equation}
In the conserving case, the chemical potential term is constant and may be
omitted. Note that, while the generalized ABC dynamics is strictly local,
the Hamiltonian is long-ranged. The fact that detailed balance is
satisfied with respect to the particle-conserving processes \eqref{eq:ABCdynamics}
and \eqref{eq:vacancyexchange} is a result of the neutrality of the
vacancies. The reason why detailed balance is also satisfied with
respect to the nonconserving processes \eqref{eq:addremoverates}
has to do with the fact that the energy $E$ of a configuration is invariant
under any translation of $ABC$ triplets, namely, $E(\ldots YABC \ldots)=E(\ldots ABCY \ldots)$,
where $Y$ stands for either a particle of any species or a
vacancy. Thus, the rates of depositing or evaporating $ABC$ triplets
could be taken as independent of the microscopic configuration in
which these processes take place, as given by the rates
\eqref{eq:addremoverates}. The change in energy corresponding to
depositing an $ABC$ triplet in a system containing $N$ particles
is given by $\Delta\tilde{\mathcal{H}}=N+1-\frac{1}{6}\left(6N+6\right)-3\mu
L=-3\mu L$. Thus, detailed balance is satisfied with respect to \eqref{eq:addremoverates}.

The conserving dynamics leads to the canonical steady state of
the Hamiltonian \eqref{eq:H_ABCgen}, while the nonconserving dynamics
leads to the grand canonical one. As a result of the long-range nature of the
interactions, the two ensembles need not be equivalent, yielding
different phase diagrams. In order to carry out this analysis we take
$q=\exp\left(-\beta/L\right)$ and turn to the continuum limit \cite{Clincy2003,Ayyer2009}.
The local concentration of $A,\, B$, and $C$ particles at the
point $x=i/L$ is represented by the density profile $\rho_{n}(x) \;
(n=A,B,C)$ with $\rho(x)=\rho_A(x)+\rho_B(x)+\rho_C(x)$. The steady-state 
distribution of the density profiles is given by
$P\left[\rho_{n}(x)\right]=\exp\left\{
-L\mathcal{F}\left[\rho_{n}(x)\right]\right\} $, where
$\mathcal{F}\left[\rho_{n}(x)\right]$ is the free energy functional.
The equilibrium profile can thus be found by minimizing the free
energy functional with respect to $\rho_{n}(x)$ under the equal
densities condition $\int_{0}^{1}\rho_{n}(x)dx\equiv r/3$.

We begin by considering the conserving dynamics (i), where the
overall density $r$ is fixed, and the free energy functional is
\begin{eqnarray}
&&\mathcal{F}[\rho_{n}(x)]=\int_{0}^{1}dx[\rho_{A}(x)\ln\rho_{A}(x)+\rho_{B}(x)\ln\rho_{B}(x)\nonumber \\
&&+\rho_{C}(x)\ln\rho_{C}(x)+(1-\rho(x))\ln(1-\rho(x))] \nonumber \\
&&+\beta\left( \int_{0}^{1}dx\int_{0}^{1-x}dz[\rho_{A}(x)\rho_{C}(x+z)\right.\nonumber \\
&&\left. +\rho_{B}(x)\rho_{A}(x+z)+\rho_{C}(x)\rho_{B}(x+z)]-\frac{1}{6}r^{2}\right) .
\label{eq:Fconserving}
\end{eqnarray}
Here the first term corresponds to the entropy, and the second term
corresponds to the energy.

At high temperatures $T\equiv 1/\beta$, the free energy is minimized by
the homogeneous density profiles $\rho_{n}(x)=r/3$, corresponding to
the disordered phase. In order to find the transition, we note that
in the ordered phase the density profiles are expected to satisfy
$\rho_{B}(x)=\rho_{A}(x-1/3)$ and $\rho_{C}=\rho_{A}(x+1/3)$.
Assuming a smooth transition to the ordered phase we expand
$\rho_{A}(x)$ close to the homogeneous solution:
\begin{equation}
\rho_{A}(x)=\frac{r}{3}+\sum_{m=1}^{\infty}a_{m}\cos(2\pi
mx).\label{eq:rhoAseries}
\end{equation}
This profile is arbitrarily chosen among all its translationally related
degenerate profiles. Substituting
\eqref{eq:rhoAseries} in Eq.\ \eqref{eq:Fconserving}, together with
the symmetry conditions on $\rho_{B}(x)$ and $\rho_{C}(x)$, results
in a series expansion for the free energy $\mathcal{F}$ in powers of
the amplitudes $a_{m}$. Instabilities of the homogeneous state are
dominated by the first mode $a_{1}$, while higher modes $a_{m}$ are
driven by it. Applying the equilibrium conditions
$\partial\mathcal{F}/\partial a_{m}=0$, one can express the
coefficients  $a_{m}$ for $m>1$ in powers of $a_{1}$. This leads to
a Landau expansion for the free energy $\mathcal{F}$, with $a_{1}$
serving as an order parameter:
\begin{equation}
\mathcal{F}[\rho_{A}(x)]=\mathcal{F}\left(\frac{r}{3}\right)+f_{2}a_{1}^{2}+f_{4}a_{1}^{4}+\ldots,
\end{equation}
where
\begin{equation}
f_{2} =  \frac{9}{4r}-\frac{3\sqrt{3}\beta}{8\pi}\,;\,f_{4} = \frac{81}{32r^{3}}\left(\frac{\sqrt{3}\beta r+6\pi}{\sqrt{3}\beta r+12\pi}\right).\label{eq:fterms}
\end{equation}
Since $f_{4}>0$ for any $r$ and $\beta$, there is a second-order
phase transition at  $f_{2}=0$, where $\beta_{c}=2\pi\sqrt{3}/r$.

\begin{figure}[t]
\noindent
\begin{centering}\includegraphics[scale=0.4]{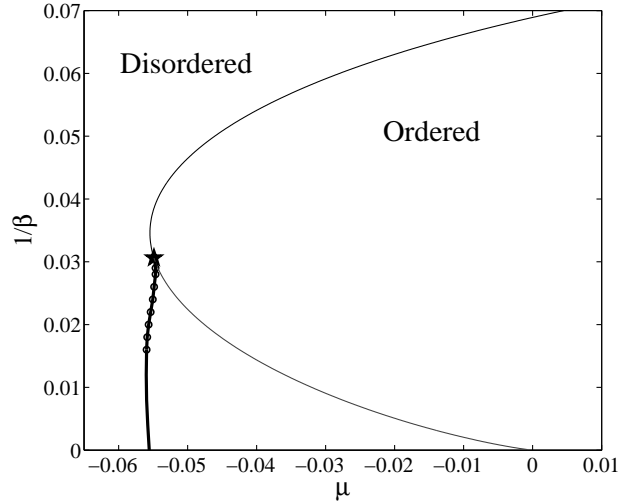}\par\end{centering}
\caption{\label{fig:PhaseDiag_gammasixth}The $(1/\beta,\mu)$ phase
diagram of the  ABC model. For conserving dynamics the model
exhibits a line of continuous transition [thin solid line, Eq.
\eqref{eq:muH}]. For nonconserving dynamics the model exhibits a
multicritical (fourth-order) point (star), beyond which the transition becomes first-order 
(thick solid line) determined numerically (circles).}
\end{figure}

By evaluating the chemical potential using
$\mu=\frac{1}{\beta}\frac{\partial}{\partial
r}\mathcal{F}\left(r/3\right)$, the critical line is obtained:
\begin{equation}
\mu=\frac{1}{\beta}\left[\ln\left(\frac{2\pi}{\sqrt{3}\beta}\right)
-\ln\left(1-\frac{2\pi\sqrt{3}}{\beta}\right)\right].\label{eq:muH}
\end{equation}
The $(1/\beta,\mu)$ phase diagram is shown in Fig. \ref{fig:PhaseDiag_gammasixth}.

In order to evaluate the phase diagram corresponding to the
nonconserving dynamics (ii), we consider the Gibbs free energy
$\mathcal{G}[\rho_{n}(x)]=\mathcal{F}\left[\rho_{n}(x)\right]-\beta\mu r$
and study the stability of the homogeneous phase at a given chemical
potential $\mu$. To this end we expand $\mathcal{G}$ in small deviations from the
homogeneous profile. Here, however, one should also allow for variations of the
overall density $\delta r$. Thus, the $A$-particle density profile close to the
transition can be written as
\begin{equation}
\rho_{A}(x)=\frac{r}{3}+\frac{\delta r}{3}+\sum_{m=1}^{\infty}a_{m}\cos(2\pi
mx).
\end{equation}
As before, the $B$ and $C$ density profiles are obtained by
translation operations. The equilibrium conditions
$\partial\mathcal{G}/\partial\left(\delta r\right)=0$ and
$\partial\mathcal{G}/\partial a_{m}=0$, for $m>1$ result in the
Landau expansion of $\mathcal{G}$ in terms of $a_{1}$,
\begin{equation}
\mathcal{G}\left[\rho_{A}(x)\right]=\mathcal{G}\left(\frac{r}{3}\right)
+g_{2}a_{1}^{2}+g_{4}a_{1}^{4}+g_{6}a_{1}^{6}+\ldots~,
\label{eq:Gexpamsion}
\end{equation}
\vspace{-0.5cm}
\begin{equation}
\hspace{-2.5cm}\mathrm{with~~~~~~~~} g_{2}=f_{2}=\frac{9}{4r}-\frac{3\sqrt{3}\beta}{8\pi}~.
\end{equation}
This coefficient vanishes at $\beta_{c}=2\pi\sqrt{3}/r$, yielding
the critical line, as long as $g_4>0$. On the critical line the fourth-order
coefficient is given by
\begin{equation}
g_{4}\left(\beta_{c}\right)=\frac{27}{32r^{3}}\left(3r-1\right)~.
\end{equation}
The transition is, thus, continuous for $r>1/3$, becoming first-order 
at a multicritical point (MCP) $r=1/3$, where $g_{2}=g_{4}=0$.
In the $\left(1/\beta,\mu\right)$ plane, the MCP is given by
\begin{equation}
\beta_{{\rm MCP}}=6\pi\sqrt{3}\;;\;
\mu_{{\rm MCP}}=-\frac{\ln 6}{6\pi\sqrt{3}}\simeq -0.0549.
\end{equation}
Calculating higher-order terms in the expansion
\eqref{eq:Gexpamsion}, we find that at the MCP
$g_6=0$ while $g_8>0$, which implies that this is a fourth-order
critical point.
\begin{figure}[here]
\noindent
\begin{centering}\includegraphics[scale=0.55]{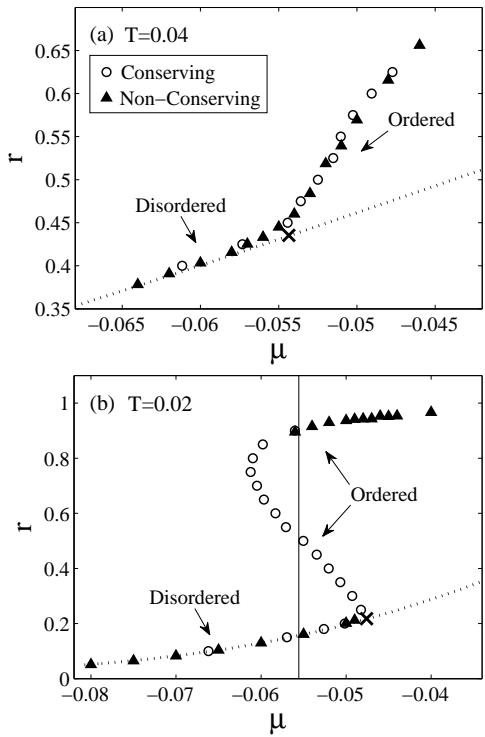}\par\end{centering}
\caption{\label{fig:rhoVsMu}The $r(\mu)$ curve obtained from Monte Carlo
simulations with $L=1800$ and temperatures (a) above and (b) below
the multicritical point. In (a) the two types of dynamics result in
the same curve with a continuous transition. The expected critical
point is marked by $\mathbf x$; deviations are due to the finite
size of the system. In (b) the conserving dynamics results in a
continuous transition ($\mathbf x$), while the nonconserving
dynamics displays a discontinuity in $r$ at a first-order
transition. Hysteretic behavior is observed. The expected transition
point obtained by minimization of $\mathcal{G}$ is indicated by the
vertical line. Dotted lines correspond to homogeneous profiles. }
\end{figure}

To complete the phase diagram we note that at $T=0$ the system is
fully phase-separated, with randomly distributed vacancies. For such
states one has $\mathcal{G}_{T=0}=-r^{2}/18-\mu r$. Minimizing
$\mathcal{G}$ with respect to $r$, one finds a first-order transition
at $\mu=-1/18$ between an empty state $(r=0)$ for $\mu<-1/18$ and a
phase-separated state with $r=1$ for $\mu>-1/18$. The first-order
line connecting the MCP and the $T=0$ transition point is found
numerically by integrating the dynamical equations
$\partial\rho_{n}(x,t)/\partial
t=-\delta\mathcal{G}/\delta\rho_{n}(x,t)$.
While these equations do not represent the actual dynamics of the
system, their steady-state solution reproduces the steady-state
profile corresponding to the minimum of $\mathcal{G}$.

The phase diagram of the nonconserving ABC model is given in Fig.
\ref{fig:PhaseDiag_gammasixth}, where it is compared with that of
the conserving model. While the transition to the phase-separated
state in the conserving model is continuous throughout the
$(1/\beta,\mu)$ plane, the transition in the nonconserving model
changes character at a fourth-order critical point. The transition
lines in the two models coincide when both are second-order. However,
the nonconserving model tends to enhance phase separation, where it
is found to exist in domains in the $(1/\beta,\mu)$ plane in which
the conserving model displays a homogeneous phase. This is a
characteristic behavior of systems with long-range interactions,
resulting from the nonadditivity of these systems
\cite{Mukamel2010,Campa2009}.

In order to verify the picture emerging from the continuum
approximation, Monte Carlo simulations were performed for (i)
conserving dynamics [rules \eqref{eq:ABCdynamics} and
\eqref{eq:vacancyexchange}] and (ii) nonconserving dynamics [rules
\eqref{eq:ABCdynamics}, \eqref{eq:vacancyexchange} and
\eqref{eq:addremoverates}], at temperatures above and below the MCP
($T_{{\rm MCP}}\simeq0.03$). In the density-conserving case, the chemical
potential $\mu$ is calculated by using the Creutz algorithm \cite{Creutz1983}.
In this method, particles may be exchanged between the system and an external single
degree of freedom, a {}``demon,'' with $N_{system}+N_{demon}=N$. The
chemical potential $\mu$ is determined by calculating the
distribution: $P\left(N_{demon}\right)\sim\exp\left(-\beta\mu
N_{demon}\right)$. The results for the density curves $r(\mu)$ are
shown in Fig. \ref{fig:rhoVsMu}. Above the MCP, both the conserving
and nonconserving dynamics yield the same curve within the
numerical accuracy. On the other hand, below $T_{{\rm MCP}}$ the
nonconserving dynamics leads to a discontinuous change in the total
density, with hysteretic behavior, consistent with a first order
phase transition. The conserving dynamics result in a continuous
transition at the expected critical point, realizing density values
which cannot be reached under nonconserving dynamics.

The canonical and grand canonical phase diagrams of the generalized
ABC model at equal densities may serve as a very useful starting point
for analyzing the phase diagram at nonequal densities, where
detailed balance is not satisfied and where a free energy cannot be
defined. Indeed, analysis of the dynamical equations in the continuum
limit and numerical simulations carried out with the density of one
species different from the other two yield a qualitatively
similar phase diagram to that of Fig. \ref{fig:PhaseDiag_gammasixth} \cite{Cohen2010}.

\acknowledgements We thank O. Cohen, S. Gupta, O. Hirschberg, Y. Kafri
and G. M. Sch$\ddot{u}$tz for helpful discussions. The
support of the Israel Science Foundation (ISF) and the Minerva Foundation with 
funding from the Federal German Ministry for Education and Research is gratefully
acknowledged.

\end{document}